 \documentclass[preprint,12pt]{elsarticle}

\usepackage{epsfig}

\usepackage{amssymb}
 \usepackage{amsthm}
 \usepackage{amsmath}
\usepackage{xcolor}  

\journal{ }

\begin{document}

\begin{frontmatter}



\title{CROSS-SECTIONS FOR THE ${^{27}\!\rm{Al}}(\gamma,\textit{x})^{24}\rm{Na}$ MULTIPARTICLE REACTION AT \mbox{$E_{\rm{\gamma max}}$ = 40 $\div$ 95 MeV}}

\author{A.N. Vodin\corref{cor1}\fnref{label2}}
\ead{vodin@kipt.kharkov.ua}
 \cortext[cor1]{Corresponding author}
 \fntext[label2]{Akademicheskaya St., 1, Kharkov, Ukraine, 61108}

\author{O.S. Deiev, I.S. Timchenko, S.N. Olejnik}
\address{National Science Center "Kharkov Institute of Physics and Technology", \\
 1 Akademicheskaya St., 61108 Kharkov, Ukraine}

\begin{abstract}
The bremsstrahlung flux-averaged cross-sections $\langle{\sigma(E_{\rm{\gamma max}})}\rangle$ and the cross-sections per equivalent photon $\langle{\sigma(E_{\rm{\gamma max}})_{\rm{Q}}}\rangle$ were measured for the photonuclear multiparticle reaction $^{27}\!\rm{Al}(\gamma,\textit{x}; \textit{x} = {^{3}\rm{He}} + pd + 2pn)^{24}\rm{Na}$ at bremsstrahlung end-point energies ranging from 40 MeV  to 95 MeV. The experiments were performed using the beam from the NSC KIPT electron linear accelerator LUE-40 with the use of the $\gamma$-activation technique. The bremsstrahlung quantum flux was calculated with the program GEANT4 and, in addition, was monitored by means of the  $^{100}\rm{Mo}(\gamma,n)^{99}\rm{Mo}$ reaction. The cross-sections $\sigma(E)$ were computed using the TALYS1.9 code with the default options. The measured average cross-sections \mbox{$\langle{\sigma(E_{\rm{\gamma max}})}\rangle$}  and $\langle{\sigma(E_{\rm{\gamma max}})_{\rm{Q}}}\rangle$ have appeared to be higher by factors of 2.0 to 2.4 than the theoretical results. The experimental results have been found to be in good agreement with the data of other laboratories. Consideration is given to special features of calculation of $\langle{\sigma(E_{\rm{\gamma max}})}\rangle$ and $\langle{\sigma(E_{\rm{\gamma max}})_{\rm{Q}}}\rangle$ for the $^{27}\!\rm{Al}(\gamma,\textit{x})^{24}\rm{Na}$ reaction, with occurrence of three $^{27}\!\rm{Al}$ photodisintegration channels. The paper also discusses the possibility of using the $^{27}\!\rm{Al}(\gamma,\textit{x})^{24}\rm{Na}$ reaction for monitoring the bremsstrahlung $\gamma$-quantum flux in the photon energy region above 30 MeV.
\end{abstract}




\end{frontmatter}

\section{Introduction}
\label{Int}
   Most of the data on the energy dependence of photonuclear reaction cross-sections were obtained from the studies of the giant dipole resonance (GDR) in the experiments performed on bremsstrahlung $\gamma$-ray/quasi-monoenergetic photon beams \cite{1,2} at $\gamma$-energies of up to $\sim$30~MeV. The investigations in this energy range were carried out mainly for the reactions with escape of one or two particles by both the direct nucleon registration method and the induced activity technique. At the same time, the research into nuclear photodisintegration in the energy range above the GDR and up to the pion production threshold ($E_{\rm{th}} \approx 145$~MeV) is of interest in connection with the change in the mechanism of photon interaction with nuclei in this energy region. It may give the key to the fundamental information on the competition of two mechanisms of nuclear photodisintegration, viz., the GDR excitation and the quasi-deuteron photoabsorption. However, the general shortage of experimental data on nuclear photodisintegration in the energy range from 30 to 145 MeV severely restricts both the general insight into the processes of $\gamma$-quantum interaction with nuclei in the energy range under discussion, and the model-approach testing capabilities \cite{3,4}.
   
      The experimental study of photonuclear reactions calls for the accurate knowledge of $\gamma$-fluxes that have interacted with the irradiated target. To this end, one makes use of the known reaction cross-sections measured with good accuracy in the corresponding $\gamma$-energy region. In particular, V.~Di~Napoli, et al. \cite{5,6} have proposed to use the reaction $^{27}\!\rm{Al}(\gamma,\textit{x})^{24}\rm{Na}$ as a monitor of the bremsstrahlung $\gamma$-flux at the energies \mbox{$E_{\rm{\gamma max}} > 30$~MeV}. In their studies the authors have revealed a linear dependence of the values of average cross-sections per equivalent photon $\langle{\sigma(E_{\rm{\gamma max}})_{\rm{Q}}}\rangle$ on the bremsstrahlung $\gamma$-ray energy in the range from 300 to 1000~MeV.  
      
    The photodisintegration of the $^{27}\!\rm{Al}$ nucleus with production of $^{24}$Na was investigated in a number of works \cite{5,6,7,8,9,10,11,12} at $E_{\rm{\gamma max}} = 20 \div 1000$~MeV. As consequence, determined were the experimental values of bremsstrahlung flux-averaged cross-sections  $\langle{\sigma(E_{\rm{\gamma max}})}\rangle$ and the cross-sections per equivalent photon $\langle{\sigma(E_{\rm{\gamma max}})_{\rm{Q}}}\rangle$. 

However, on photodisintegration of the $^{27}\!\rm{Al}$ nucleus at $E_{\rm{\gamma max}} > 30$~MeV, the final product nucleus $^{24}$Na can be produced simultaneously in several reaction channels having different energy thresholds $E_{\rm{th}}$. This must be taken into account in the experiments with the use of the induced activity method, wherein the $\gamma$-radiation of the final product nucleus is registered. So, in the thus far used representation of the experimental data in the form of the average cross-section $\langle{\sigma(E_{\rm{\gamma max}})}\rangle$, the reaction threshold energy value $E_{\rm{th}}$ should be used in the calculations. It is evident that the use of different $E_{\rm{th}}$ values will result in different values of $\langle{\sigma(E_{\rm{\gamma max}})}\rangle$. This is particularly essential when determining the average cross-section in the vicinity of the minimal threshold of the $^{27}\!\rm{Al}(\gamma,\textit{x})^{24}\rm{Na}$ reaction under consideration.

   The present work is concerned with the measurements of average cross-sections $\langle{\sigma(E_{\rm{\gamma max}})}\rangle$ and $\langle{\sigma(E_{\rm{\gamma max}})_{\rm{Q}}}\rangle$ for the multiparticle reaction $^{27}\!\rm{Al}(\gamma,\textit{x})^{24}\rm{Na}$ in the energy range $E_{\rm{\gamma max}} = 40 \div 95$~MeV with the step $\triangle E_{\rm{\gamma max}} = 2 \div 4$~MeV. Consideration is given to special features of calculating these average cross-section values. Note that the notation $(\gamma,\textit{x})$ implies that $\textit{x} = {^3\rm{He}} + \rm{pd} + 2pn$, and thus we investigate the total average cross-sections $\langle{\sigma(E_{\rm{\gamma max}})}\rangle$ and $\langle{\sigma(E_{\rm{\gamma max}})_{\rm{Q}}}\rangle$ of the mentioned reaction. In this context, to consider the cross-section for the $^{27}\!\rm{Al}(\gamma,2pn)^{24}\rm{Na}$ reaction as the total one is not exactly correct, even though this channel is dominant in comparison with other reaction channels observed in the photodisintegration of $^{27}\!\rm{Al}$ with the production of the final product nucleus $^{24}$Na. 

\section{Experimental procedure}
\label{Exp proced}
\begin{figure}
   	\center{\includegraphics[scale=0.4]{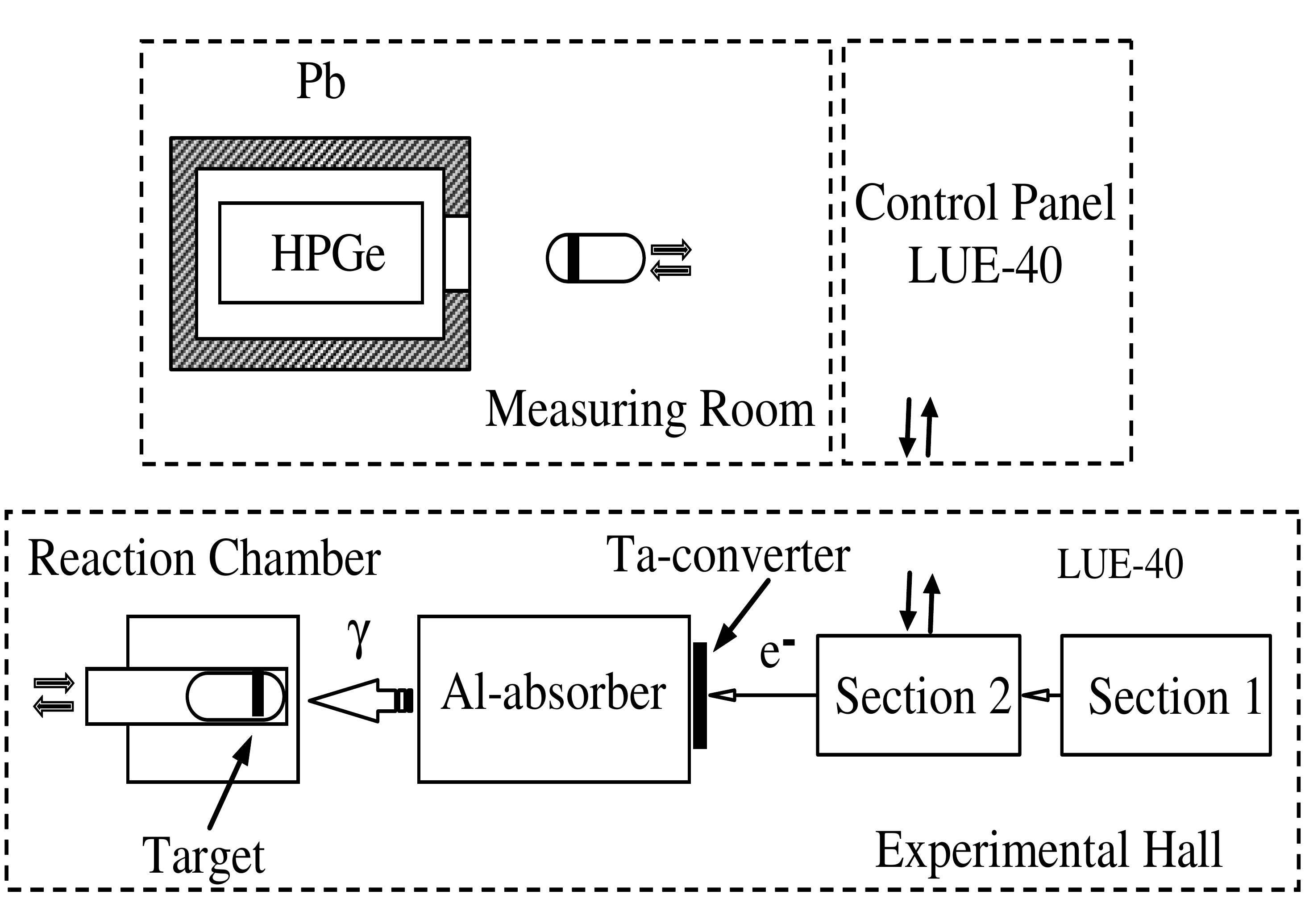}}
   	\caption{Experimental schematic diagram including three units shown with a dashed line.  Above -- the measuring room and the control panel of the LUE-40 accelerator, below -- the experimental hall.}
   	\label{fig1}
   \end{figure}

   The experiments were performed using the bremsstrahlung $\gamma$-beam from the NSC KIPT electron linear accelerator LUE-40 using the method of induced activity of the final product nucleus of the reaction. The schematic of the experiment is shown in Fig.~\ref{fig1} \cite{13,14}.
   
      The accelerator parameters make it possible to vary the energy of accelerated electrons in the range from 30 to 100~MeV at the average beam current $I_e = 3~\mu$A. In this case, the electron energy spectrum width at FWHM makes $\triangle E_e/E_e \sim 1 \div 1.5 \%$ at a pulse-repetition frequency of 50~Hz and the pulse length of $10~\mu$s. 
      
         The bremsstrahlung $\gamma$-radiation was generated on passing a pulsed electron beam through a tantalum metal plate, 1.05~mm in thickness (the radiation length of Ta is $\sim\!4.1$~mm). The Ta-converter was fixed on the aluminum cylinder, 100 mm in diameter and 100 or 150 mm in thickness. The aluminum cylinder was used to absorb the electrons that have passed through the converter. The usage of the Al-absorber, which fulfilled the functions of both the additional converter and the scatterer of the primary $\gamma$-radiation, caused some distortion of the bremsstrahlung spectrum (mainly in its low energy part). However, these $\gamma$-flux distortions were fully taken into account when using the GEANT4 code for calculation purposes, similar to refs.~\cite{13,14}. In addition, the bremsstrahlung $\gamma$-flux was also estimated from the $^{100}\rm{Mo}(\gamma,n)^{99}\rm{Mo}$ reaction yield by using the approach of ref.~\cite{14}. 

   For the experiments, the $^{\rm{nat}}$Mo and $^{27}\!\rm{Al}$ targets were prepared, which represented thin discs of the like diameters 8~mm and the thicknesses of 0.1~mm for molybdenum and 1~mm for aluminum, that corresponding to the masses $m \approx$ 60 and 135 mg, respectively. The both targets were simultaneously exposed to bremsstrahlung $\gamma$-quanta for the time $t_{\rm{irr}}$ = 30~min at all electron energy values. The targets were delivered to the reaction chamber in a special aluminum capsule by means of the pneumatic conveyor system. After the exposure, the both targets were transferred to the measuring room, where the induced $\gamma$-activity spectra of the samples were registered in turn by means of the HPGe detector. The measurement time was $t_{\rm{meas}}$ = 30 min per sample. The resolution of the HPGe detector Canberra GS-2018 was 1.8~keV (at FWHM) for the $E_{\gamma} = 1332$~keV $\gamma$-line of $^{60}$Co, and its efficiency was 20\% relative to the NaI(Tl) detector, 300~mm in diameter and 300~mm in thickness. The standard $\gamma$-radiation sources $^{22}$Na, $^{60}$Co, $^{133}$Ba, $^{137}$Cs, $^{152}$Eu and $^{241}\!\rm{Am}$ were used for energy/efficiency calibration of the spectrometry channel. In $\gamma$-ray spectrum measurements, the dead time of the spectrometry channel was no more than 3 to 5\%, and was determined through choosing the appropriate distance between the irradiated sample and the HPGe detector. The $\gamma$-ray spectra were analyzed using the Canberra software and the ORIGIN2015 code. 
   
      The $^{24}$Na $\gamma$-activity studies of the $^{27}\!\rm{Al}(\gamma,\textit{x})^{24}\rm{Na}$  reaction were performed through the use of the $E_{\gamma} = 1368.63$  and 2754.03~keV $\gamma$-lines. The characteristics of these $\gamma$-lines, and also, of the $\gamma$-lines observed in the decay of the product nucleus $^{99}$Mo, which was produced in the $^{100}\rm{Mo}(\gamma,n)^{99}\rm{Mo}$  reaction, are given in Table~\ref{tab1}. The difference between the $^{27}\!\rm{Al}(\gamma,{^3\rm{He}})^{24}\rm{Na}$ and $^{27}\!\rm{Al}(\gamma,2pn)^{24}\rm{Na}$ reaction thresholds reaches 7.7~MeV, this being quite essential when estimating the bremsstrahlung $\gamma$-flux value, which is used in the calculation of the total average cross-section value, $\langle{\sigma(E_{\rm{\gamma max}})}\rangle$.
      
 \begin{table}[h]
      	\caption{\label{tab1} Spectroscopic data for the product nuclei from the reactions $^{27}\!\rm{Al}(\gamma,\textit{x})^{24}\rm{Na}$ and $^{100}\rm{Mo}(\gamma,n)^{99}\rm{Mo}$  \cite{15}}
      	\centering
      	\begin{tabular}{ccccc}
      		\hline	\vspace{1ex}
      		Nuclear reaction & $E_{\rm{th}}$,~MeV & $T_{1/2}$ & $E_{\gamma}$,~keV & $I_{\gamma}$, \% \\ 	\hline
      	  \vspace{1ex}	   
      \begin{tabular}{c}$^{27}\!\rm{Al}(\gamma,{^3\rm{He}})^{24}\rm{Na}$\\ $^{27}\!\rm{Al}(\gamma,pd)^{24}\rm{Na}$ \\ $^{27}\!\rm{Al}(\gamma,2pn)^{24}\rm{Na}$ 
      \end{tabular}& 
  \begin{tabular}{c}23.710\\ 29.203 \\31.428  
  \end{tabular}&       	  $14.9590 \pm 0.0012$~h & 
      	 \begin{tabular}{c}1368.633\\2754.028 
      	 \end{tabular}& 
           	 \begin{tabular}{c}$100$\\$99.944 \pm 0.004$
     \end{tabular}\\   
$^{100}\rm{Mo}(\gamma,n)^{99}\rm{Mo}$ & 8.29 & $65.94 \pm 0.01$~h & 
      		\begin{tabular}{c}140.51\\739.50 
      		\end{tabular}& 
      	\begin{tabular}{c}$89.43 \pm 0.23$\\$12.13 \pm 0.12$ 
      	\end{tabular} \\	
         \hline
      	\end{tabular}	        
  \end{table}

     The accuracy of bremsstrahlung flux-averaged cross-section measurements was determined as a quadratic sum of statistical and systematic errors. The statistical error in the observed $\gamma$-activity is mainly due to the statistics calculation, and is estimated to be 1 to 3\%. The error is dependent on the $\gamma$-line intensity and the background conditions of spectral measurements. The systematic errors are due to the uncertainties about: 1) the exposure time (0.5\%); 2) the electron current (0.5\%); 3) the $\gamma$-rays registration efficiency ($\sim$3\%), which is mainly attributed to $\gamma$-radiation source uncertainties; 4) normalization of experimental data to the yield of the monitoring reaction $^{100}\rm{Mo}(\gamma,n)^{99}\rm{Mo}$ ($0.5 \div 2\%$); 5) the GEANT4 miscalculation of the quantum flux ($\sim$1.5\%). Thus, the experimental error of the obtained data ranges within 5 to 6\%. 
      
      \section{Calculation of average cross-sections for the $^{27}\!\rm{Al}(\gamma,\textit{x})^{24}\rm{Na}$ reaction}
      \label{Calc of ave}   
  
     The total and partial cross-sections $\sigma(E)$ for the reaction $^{27}\!\rm{Al}(\gamma,\textit{x})^{24}\rm{Na}$ were calculated for monochromatic photons in the TALYS1.9 code with the default options \cite{16}. The cross-sections $\sigma(E)$ were averaged over the bremsstrahlung flux $W(E,E_{\rm{\gamma max}})$ in the energy range from the threshold of the corresponding reaction, $E_{\rm{th}}$, to the maximum $\gamma$-quantum energy $E_{\rm{\gamma max}}=40 \div 95$~MeV. The quantum flux was computed in GEANT4 \cite{17}. As a result, the average cross-sections $\langle{\sigma(E_{\rm{\gamma max}})}\rangle$ were obtained as: 
            
\begin{equation}\label{form1}
\langle{\sigma(E_{\rm{\gamma max}})}\rangle = \frac
{\int\limits_{E_{\rm{th}}}^{E_{\rm{\gamma max}}}\sigma(E)\cdot W(E,E_{\rm{\gamma max}})dE}
{\int\limits_{E_{\rm{th}}}^{E_{\rm{\gamma max}}}W(E,E_{\rm{\gamma max}})dE}.
\end{equation}

The calculated in this way $\langle{\sigma(E_{\rm{\gamma max}})}\rangle$ values were compared with the experimental measured average cross-sections determined by the expression

\begin{equation}\begin{split}
\langle{\sigma(E_{\rm{\gamma max}})}\rangle = \\
\frac{\lambda \triangle A}{N_x I_{\gamma} \ \varepsilon \Phi (1-\exp(-\lambda t_{\rm{irr}}))\exp(-\lambda t_{\rm{cool}})(1-\exp(-\lambda t_{\rm{meas}}))},
\label{form2}
\end{split}\end{equation}
where $\triangle A$ is the number of counts of $\gamma$-quanta in the full absorption peak (for the $\gamma$-line of the investigated reaction), $\Phi$  is the sum of bremsstrahlung quanta in the energy range from the reaction threshold $E_{\rm{th}}$ up to $E_{\rm{\gamma max}}$, $N_x$ is the number of target atoms, $I_{\gamma}$ is the absolute intensity of the analyzed $\gamma$-quanta, $\varepsilon$ is the absolute detection efficiency for the analyzed $\gamma$-quanta energy, $\lambda$ is the decay constant \mbox{($\rm{ln}2/\textit{T}_{1/2}$)}; $t_{\rm{irr}}$, $t_{\rm{cool}}$ and $t_{\rm{meas}}$ are the irradiation time, cooling time and measurement time, respectively. As is evident from eqs. (\ref{form1}) and (\ref{form2}), the average cross-section value, $\langle{\sigma(E_{\rm{\gamma max}})}\rangle$, is dependent on the bremsstrahlung energy distribution and on the reaction threshold energy $E_{\rm{th}}$.

\begin{figure}[h]
	\center{\includegraphics[scale=0.4]{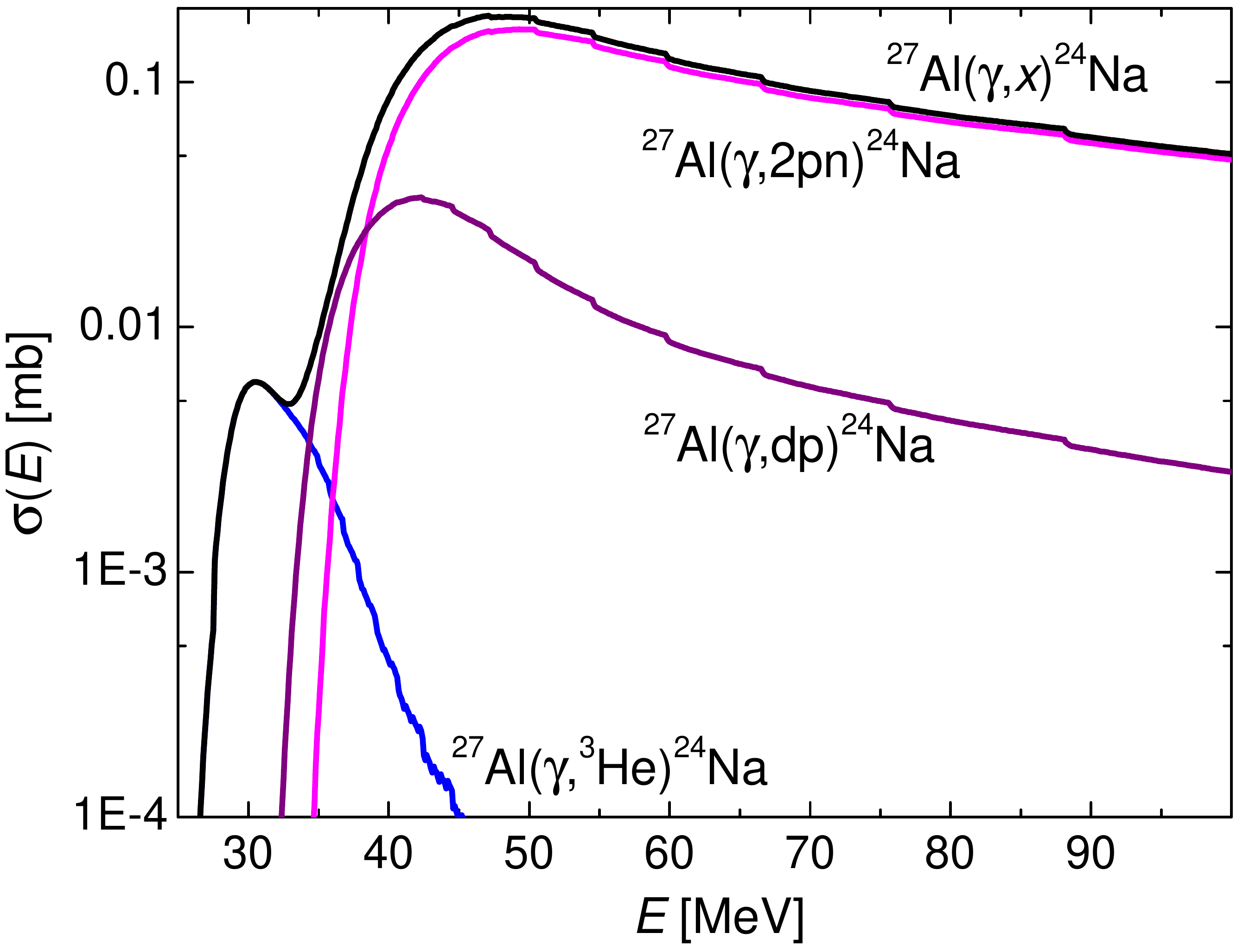}}
	\caption{TALYS1.9 computation of cross-sections $\sigma(E)$. The total $^{27}\!\rm{Al}(\gamma,\textit{x})^{24}\rm{Na}$ reaction cross-section was calculated as a sum of partial cross-sections.}
	\label{fig2}
\end{figure}

\begin{figure}[h]
	\center{\includegraphics[scale=0.4]{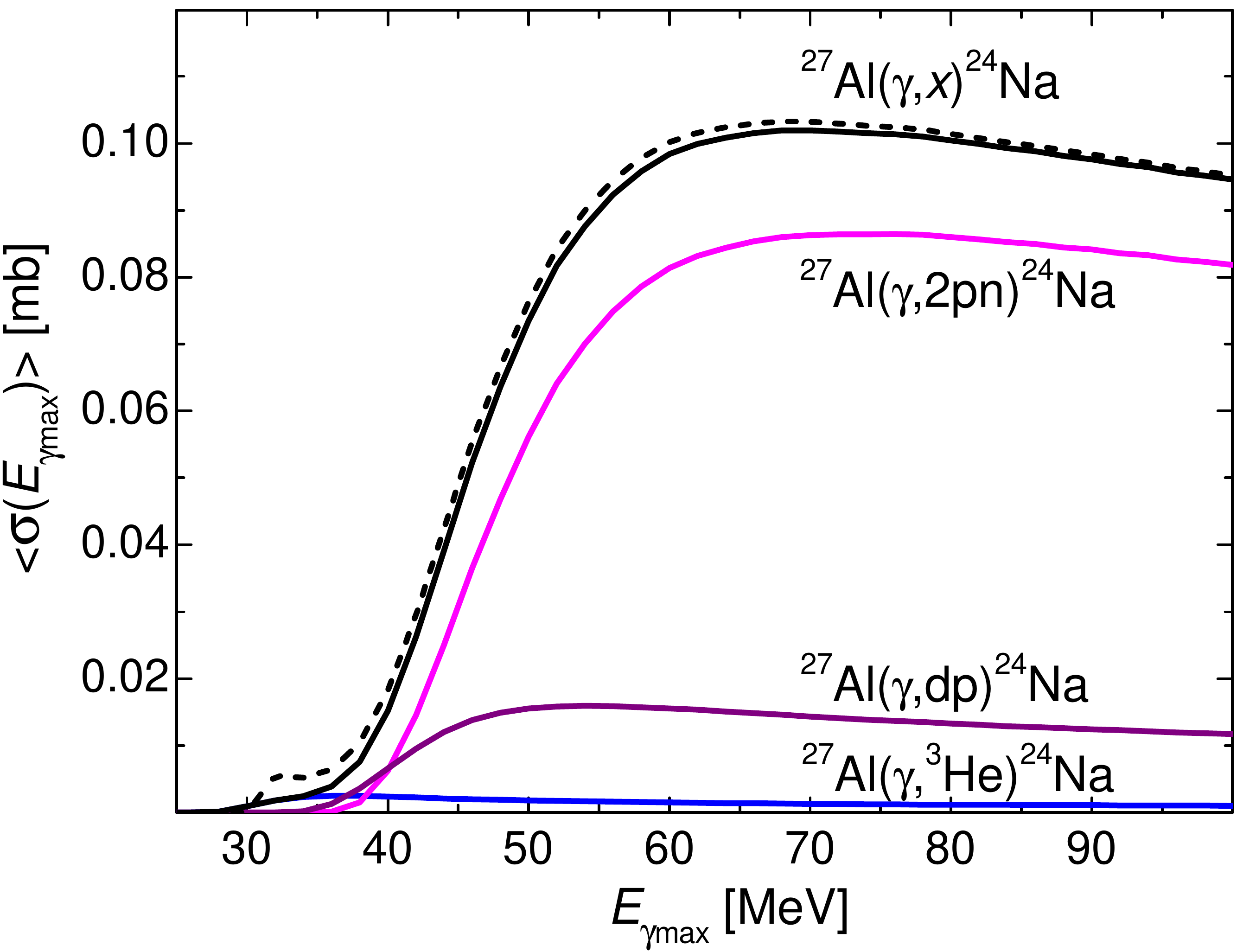}}
	\caption{Bremsstrahlung flux-averaged cross-sections $\langle{\sigma(E_{\rm{\gamma max}})}\rangle$ for $^{27}\!\rm{Al}(\gamma,\textit{x})^{24}\rm{Na}$, computed with the codes TALYS1.9 and GEANT4. The total average cross-section is shown as a sum of average partial cross-sections (solid black curve) and the total cross-section $\sigma(E)$ averaged for $E_{\rm{th}} = 31.4$~MeV (dashed black curve). The bremsstrahlung quantum flux from a thin (50~$\mu$m) Ta-converter was used for averaging.}
	\label{fig3}
\end{figure}

As noted above, $^{24}\rm{Na}$ can be produced in three $^{27}\!\rm{Al}(\gamma,\textit{x})^{24}\rm{Na}$ reaction channels, viz., $(\gamma,{^3\rm{He}})$, $(\gamma,\rm{pd})$ or $(\gamma,2\rm{pn})$. To calculate the total average cross-section for the reaction under discussion, it will be necessary to summate the partial average cross-sections, each of which is calculated with its threshold $E_{\rm{th}}$ (see Table~\ref{tab1}). Figures \ref{fig2} and \ref{fig3} show the cross-sections $\sigma(E)$ for each $^{24}\rm{Na}$ production channel in the $^{27}\!\rm{Al}(\gamma,\textit{x})^{24}\rm{Na}$ and the  respective average cross-sections calculated by eq.~(\ref{form1}) with the use of the codes TALYS1.9 and GEANT4. The two figures clearly demonstrate that at energies up to 35~MeV the cross-sections for the $(\gamma,\rm{pd})$ and $(\gamma,{^3\rm{He}})$ reactions prevail, whereas in the energy region above 50~MeV it is the $^{27}\!\rm{Al}(\gamma,2\rm{pn})^{24}\rm{Na}$ that becomes dominant. 


The total bremsstrahlung flux-averaged cross-section $\langle{\sigma(E_{\rm{\gamma max}})}\rangle$ must be calculated as a sum of partial cross-sections, whereas the calculation of $\langle{\sigma(E_{\rm{\gamma max}})}\rangle$ with the use of the total $\sigma(E)$ values would be incorrect, because in this case different reaction thresholds $E_{\rm{th}}$ for each partial cross-section must be substituted into eq.~(\ref{form1}). For example, the use of $E_{\rm{th}} = 31.4$~MeV (the threshold value of the $^{27}\!\rm{Al}(\gamma,2\rm{pn})^{24}\rm{Na}$ reaction, which is dominant in the $^{24}\rm{Na}$ production) for calculating $\langle{\sigma(E_{\rm{\gamma max}})}\rangle$ always gives a somewhat overestimated value in comparison with the total average cross-section value $\langle{\sigma(E_{\rm{\gamma max}})}\rangle$ calculated as a sum of partial average cross-sections, each with its peculiar threshold $E_{\rm{th}}$ (see solid and dashed curves in Fig.~\ref{fig3}). In case of the reaction $^{27}\!\rm{Al}(\gamma,\textit{x})^{24}\rm{Na}$, from the energy of 60~MeV onwards, the difference between the two variants of calculating $\langle{\sigma(E_{\rm{\gamma max}})}\rangle$ does not exceed 2\%, and it slowly drops off down to 0.6\% at 100~MeV.  However, in the energy range from 32 to 45~MeV this difference is considerable. The dependence of $\langle{\sigma(E_{\rm{\gamma max}})}\rangle$ on the reaction threshold makes the experimental determination procedure ambiguous for the reaction cross-sections. 

For representing experimental data on photonuclear reactions, use is also made of the average cross-section per equivalent quantum $\langle{\sigma(E_{\rm{\gamma max}})_{\rm{Q}}}\rangle$, which is determined as:

\begin{equation}\label{form3}
\langle{\sigma(E_{\rm{\gamma max}})}_{\rm{Q}}\rangle = E_{\rm{\gamma max}}\frac
{\int\limits_{0}^{E_{\rm{\gamma max}}}\sigma(E) \cdot W(E,E_{\rm{\gamma max}})dE}
{\int\limits_{0}^{E_{\rm{\gamma max}}}E\cdot W(E,E_{\rm{\gamma max}})dE}.
\end{equation}

The equation for $\langle{\sigma(E_{\rm{\gamma max}})_{\rm{Q}}}\rangle$ is written in a more complicated form than for $\langle{\sigma(E_{\rm{\gamma max}})}\rangle$. The energy dependence of $\langle{\sigma(E_{\rm{\gamma max}})_{\rm{Q}}}\rangle$ is different from that of the average cross-section $\langle{\sigma(E_{\rm{\gamma max}})}\rangle$. However, $\langle{\sigma(E_{\rm{\gamma max}})_{\rm{Q}}}\rangle$ is formally independent of the reaction threshold, this being of importance for the reactions going in several channels. This allows us to determine, the experimental $\langle{\sigma(E_{\rm{\gamma max}})_{\rm{Q}}}\rangle$ values without  considering $E_{\rm{th}}$. 

The comparison between two types of average cross-sections shows the preference of using $\langle{\sigma(E_{\rm{\gamma max}})_{\rm{Q}}}\rangle$ in the case, where several reaction channels become open at a certain $E_{\rm{\gamma max}}$ value, while the average cross-sections \mbox{$\langle{\sigma(E_{\rm{\gamma max}})}\rangle$} are more convenient in use for the reactions with one threshold $E_{\rm{th}}$. Generally, the first variant takes place if there are charged particles present  in the yield channel of the reactions of type $(\gamma,y\rm{p}\textit{x}\rm{n}; \textit{y}, \textit{x} = 1, 2, ...)$, and the second variant is peculiar to $(\gamma,x\rm{n})$.

  \section{Results and discussion}
  \label{Res_disc}
  
\begin{figure}
	\center{\includegraphics[scale=0.4]{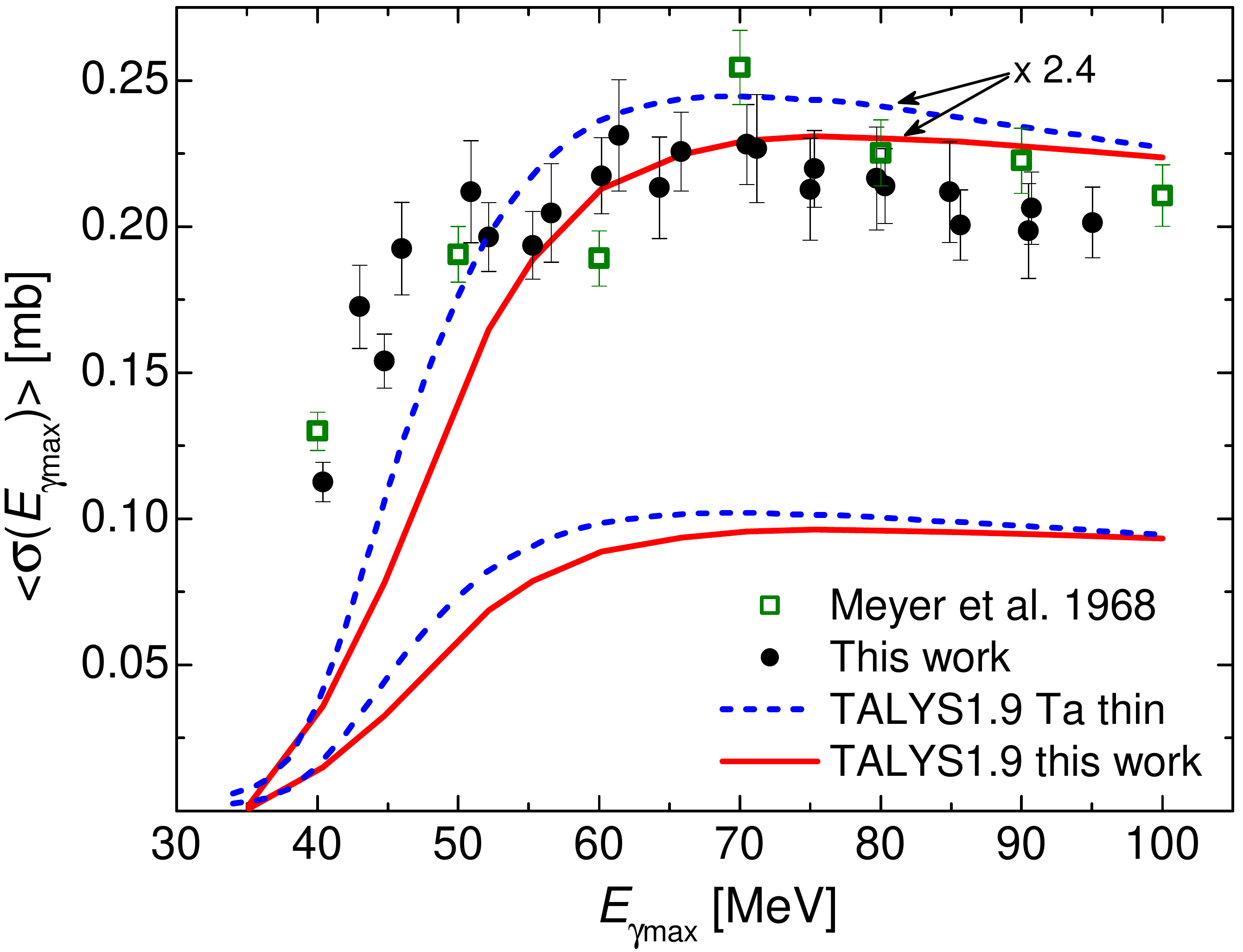}}
	\caption{Total average cross-sections $\langle{\sigma(E_{\rm{\gamma max}})}\rangle$ for the $^{27}\!\rm{Al}(\gamma,\textit{x})^{24}\rm{Na}$ reaction: data of \cite{8} -- squares, data of this work -- circles. Computations with TALYS1.9 and GEANT4: red curves -- for the present experiment conditions, blue curves -- for the thin Ta-converter, 50~$\mu$m thick. The upper line pair was obtained through multiplication of the lower line pair by a factor of 2.4.}
	\label{fig4}
	\vspace{2ex}
\end{figure}

  Figure~\ref{fig4} shows the data on the total average cross-section $\langle{\sigma(E_{\rm{\gamma max}})}\rangle$ for the reaction $^{27}\!\rm{Al}(\gamma,\textit{x})^{24}\rm{Na}$, both the measured (circles and squares) and calculated using the TALYS1.9 and GEANT4 codes (curves). The total average cross-section is shown as a sum of average partial cross-sections (see Fig.~\ref{fig3}).
The experimentally obtained $\langle{\sigma(E_{\rm{\gamma max}})}\rangle$ values were compared with the data of ref.~\cite{8}, where the result was represented in the form of average cross-sections $\langle{\sigma(E_{\rm{\gamma max}})_{\rm{Q}}}\rangle$. Note that Meyer et al. (ref.~\cite{8}) employed a thin converter, and that was taken into account in normalizing those values to the average cross-section $\langle{\sigma(E_{\rm{\gamma max}})}\rangle$ values. As is evident from Fig.~\ref{fig4}, there is fair agreement between our experimental results and the data from ~\cite{8}.

Figure~\ref{fig4} also shows the calculated total average cross-sections for two types of converters: the blue line for a thin Ta-converter, 50~$\mu$m in thickness; the red line represents the real geometry of the experiment with the Ta-converter and Al-absorber, 1.05~mm and 100~mm thick, respectively. It is obvious that the calculated average cross-section values are dependent on the bremsstrahlung spectrum shape. Hence, it follows that the experimental data obtained by averaging over the bremsstrahlung $\gamma$-quantum flux (both average cross-sections and the average cross-sections per equivalent photon) should be compared with the calculations performed for the identical bremsstrahlung $\gamma$-quantum flux, i.e., the experimental and computational geometries must be in complete agreement.

The calculation of $\langle{\sigma(E_{\rm{\gamma max}})}\rangle$ values for the $^{27}\!\rm{Al}(\gamma,\textit{x})^{24}\rm{Na}$ reaction using the TALYS1.9 code data is inconsistent with our experimental data in both the magnitude and the energy dependence form. For example, the usage of the 2.4-factor makes it possible to make into agreement the calculation and the experiment in the energy region $E_{\rm{\gamma max}}= 55 \div 70$~MeV. However, in the foreground of the energy dependence the experimental values are systematically found to be above the calculated data, and from $E_{\rm{\gamma max}} = 70$~MeV and on, the experimental values systematically occur to be below the calculations. Note that the same tendency is also observed for the measured data of ref.~\cite{8} and the corresponding calculations.

\begin{figure}
	\center{\includegraphics[scale=0.4]{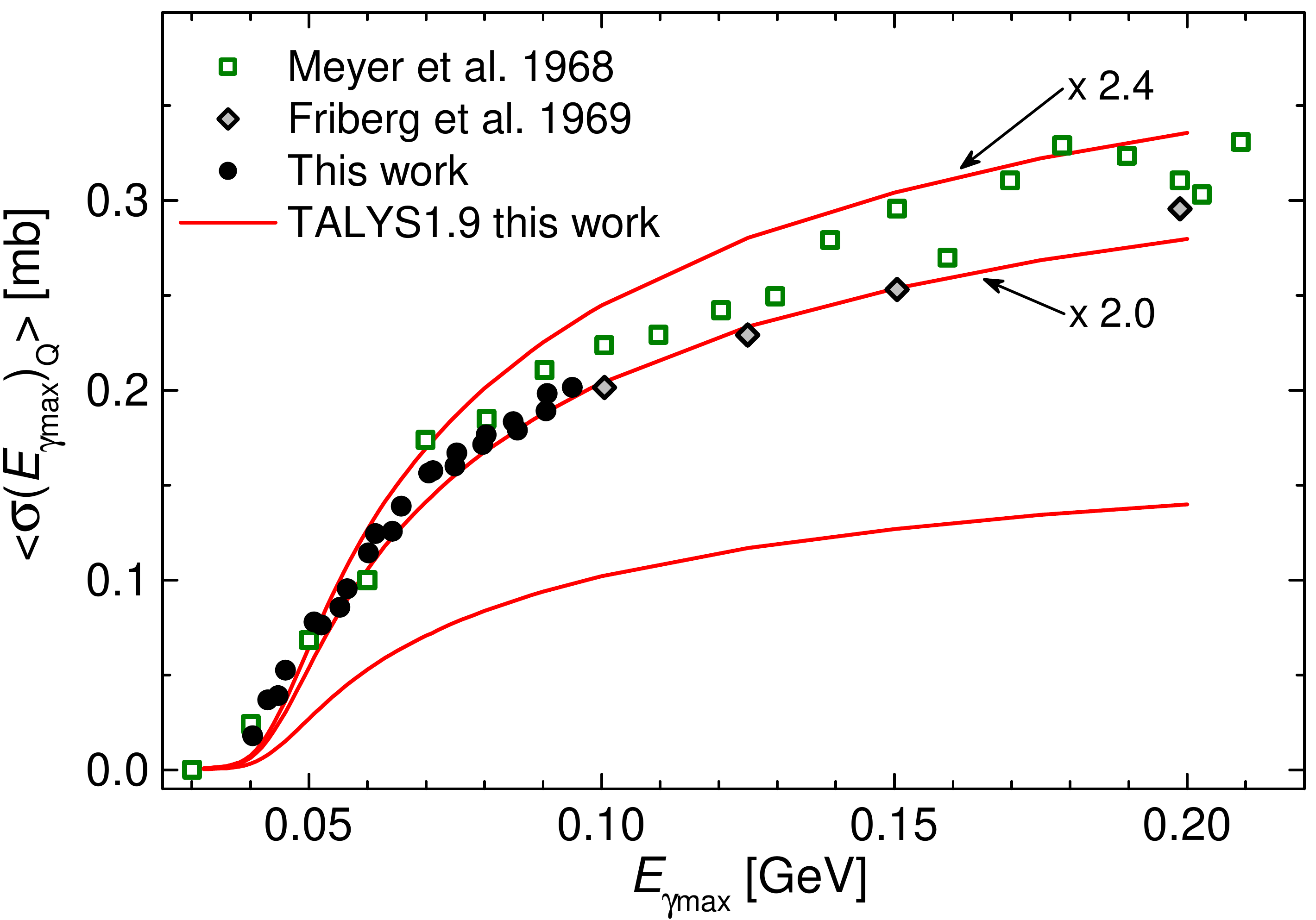}}
	\caption{Experimental total average cross-section $\langle{\sigma(E_{\rm{\gamma max}})_{\rm{Q}}}\rangle$ values for the $^{27}\!\rm{Al}(\gamma,\textit{x})^{24}\rm{Na}$ reaction, and the data computed with the TALYS1.9 and GEANT4 codes for the present experimental conditions (shown in red lines).  Experimental results: squares -- data of \cite{8}, diamonds -- \cite{11}, circles -- data of this work. Lower curve -- calculation in TALYS1.9, middle -- calculation with multiplication by 2.0, upper curve -- calculation with multiplication by 2.4. The  TALYS1.9 code is restricted to cross-section calculations up to 200 MeV.}
	\label{fig5}
\end{figure}

The most part of the experimental data available in the literature on the $^{27}\!\rm{Al}(\gamma,\textit{x})^{24}\rm{Na}$ reaction has been represented in terms of the average cross-section per equivalent photon \cite{5,6,11,12}. For comparison with those data, the experimental results of the present work are reported in terms of $\langle{\sigma(E_{\rm{\gamma max}})_{\rm{Q}}}\rangle$ according to eq.~(\ref{form3}). Figure~\ref{fig5} shows the experimental total average cross-section $\langle{\sigma(E_{\rm{\gamma max}})_{\rm{Q}}}\rangle$ values for the $^{27}\!\rm{Al}(\gamma,\textit{x})^{24}\rm{Na}$ reaction, and the calculation data obtained with the TALYS1.9 and GEANT4 codes.

As is seen from Fig.~\ref{fig5}, our experimental $\langle{\sigma(E_{\rm{\gamma max}})_{\rm{Q}}}\rangle$ values show good agreement with the data of other laboratories, in particular, with the results of refs.~\cite{5,11}. To avoid overloading the figure, the experimental errors ($\sim$6\%) are not shown.

As in the case of the average cross-section discussed above, the computation of $\langle{\sigma(E_{\rm{\gamma max}})_{\rm{Q}}}\rangle$ gives underestimated values in comparison with the experiment. Nearly all the experimental data on $\langle{\sigma(E_{\rm{\gamma max}})_{\rm{Q}}}\rangle$ at $E_{\rm{\gamma max}}$ above 50~MeV fall within the region bounded by two computation lines, viz., with multiplication of the cross-section under consideration  by the factors of 2.0 and 2.4. However, for $E_{\rm{\gamma max}} < $ 50~MeV the calculated energy dependence $\langle{\sigma(E_{\rm{\gamma max}})_{\rm{Q}}}\rangle$ is lower than the experimental result (as in the case of $\langle{\sigma(E_{\rm{\gamma max}})}\rangle$, see Fig.~\ref{fig4}). It is possible that this difference may be due to the TALYS1.9 underestimation of the cross-section at energies up to 40~MeV.

\begin{figure}
	\center{\includegraphics[scale=0.4]{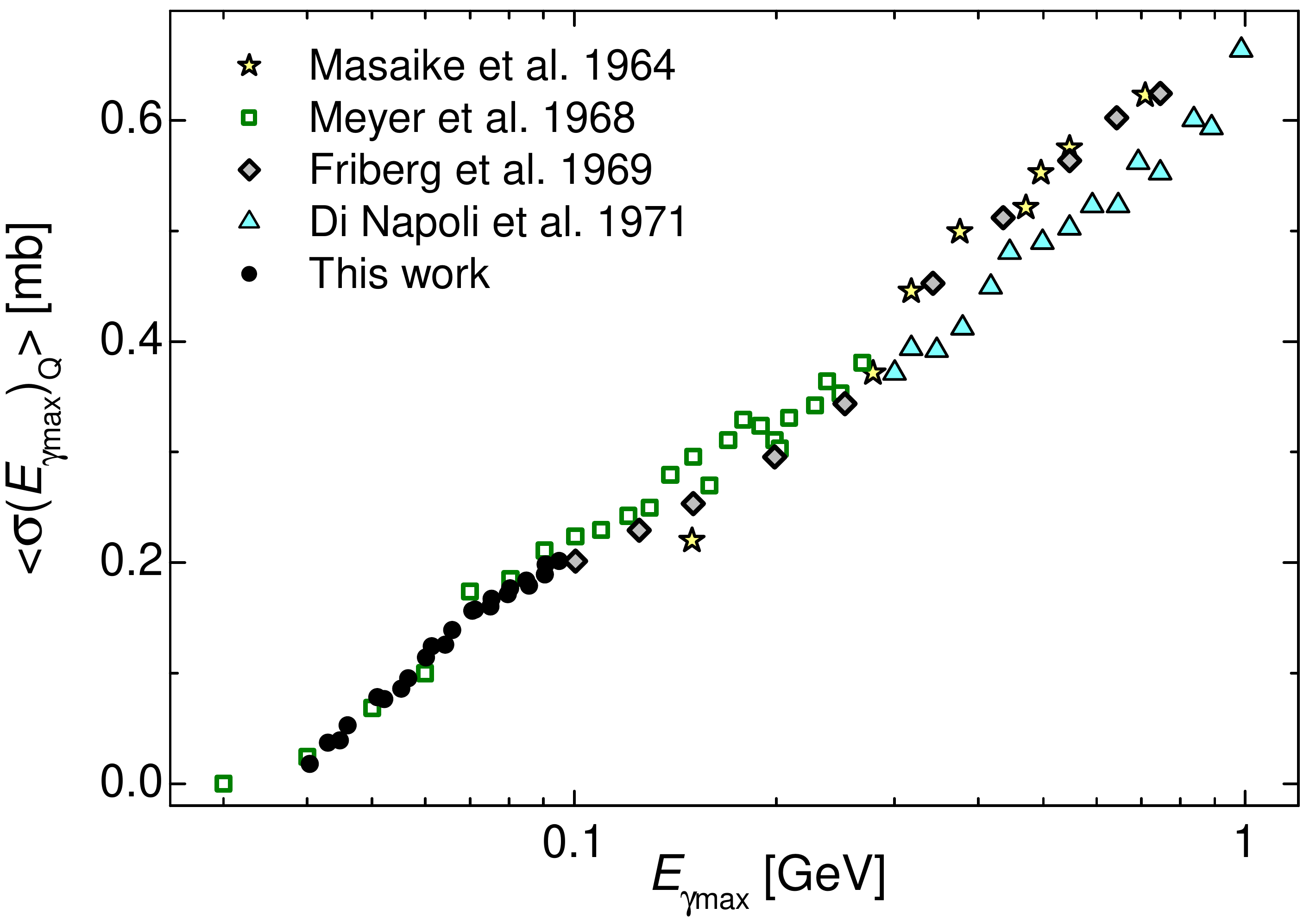}}
	\caption{Experimental data on the total average cross-sections $\langle{\sigma(E_{\rm{\gamma max}})_{\rm{Q}}}\rangle$  per equivalent photon for the  $^{27}\!\rm{Al}(\gamma,\textit{x})^{24}\rm{Na}$ reaction: stars -- \cite{12},  squares -- \cite{8}, diamonds -- \cite{11}, triangles -- \cite{5,6}, circles -- data of this work.}
	\label{fig6}
\end{figure}

Figure~\ref{fig6} illustrates the experimental data on $\langle{\sigma(E_{\rm{\gamma max}})_{\rm{Q}}}\rangle$, obtained by different laboratories \cite{5,6,8,11,12} in a wide range of $\gamma$-quantum energies. Their analysis shows that on the semi-logarithmic scale the energy dependence of $\langle{\sigma(E_{\rm{\gamma max}})_{\rm{Q}}}\rangle$ is close to the linear function. This fact makes the $^{27}\!\rm{Al}(\gamma,\textit{x})^{24}\rm{Na}$ reaction convenient for monitoring the bremsstrahlung  $\gamma$-quantum flux at the energies $E_{\rm{\gamma max}} > $ 30~MeV.

It can be seen in Fig.~\ref{fig6} that the data of refs.~\cite{5,6,8,11,12} differ in both the magnitude and the behavior of the energy dependence of the cross section $\langle{\sigma(E_{\rm{\gamma max}})_{\rm{Q}}}\rangle$. At $E_{\rm{\gamma max}} = $ 200 to 300~MeV, one might expect an insignificant kink of the energy dependence curve that can be attributed to the presence of the second maximum in the $^{27}\!\rm{Al}(\gamma,\textit{x})^{24}\rm{Na}$ cross-section at $\sim$500~MeV \cite{5,9}.  However, the general pattern looks more likely as the linear dependence of $\langle{\sigma(E_{\rm{\gamma max}})_{\rm{Q}}}\rangle$ on the energy, as expected by the authors of ref.~\cite{6}.

\section{Conclusions}
\label{Concl}
The present work has been concerned with the measurements of total average cross-sections $\langle{\sigma(E_{\rm{\gamma max}})}\rangle$ and  $\langle{\sigma(E_{\rm{\gamma max}})_{\rm{Q}}}\rangle$ for the multiparticle reaction $^{27}\!\rm{Al}(\gamma,\textit{x})^{24}\rm{Na}$  at energies ranging from 40~MeV  to 95~MeV (the step being $\triangle E_{\rm{\gamma max}} = 2 \div 4$~MeV). The induced activity technique was used for the measurements. The bremsstrahlung $\gamma$-quantum flux was computed with the GEANT4 code, and in addition, was monitored by means of the  $^{100}\rm{Mo}(\gamma,n)^{99}\rm{Mo}$ reaction. The obtained $\langle{\sigma(E_{\rm{\gamma max}})}\rangle$ and  $\langle{\sigma(E_{\rm{\gamma max}})_{\rm{Q}}}\rangle$ values are in good agreement with the data of other laboratories. 

It has been found that the calculation using the TALYS1.9 code with the default options gives the average cross-section values lower by factors of $\sim$2.0 to 2.4 in comparison with the values obtained by experiment. Besides, in the energy range under study the behavior of the average cross-sections $\langle{\sigma(E_{\rm{\gamma max}})}\rangle$ and  $\langle{\sigma(E_{\rm{\gamma max}})_{\rm{Q}}}\rangle$ is somewhat different from that predicted by calculations. It has been demonstrated that the experimental data should be compared with the calculated values for the identical bremsstrahlung quantum flux, i.e., the experimental and computational geometries are to be in complete agreement. 

The comparison between two types of average cross-sections has revealed the advantage of using the cross-section $\langle{\sigma(E_{\rm{\gamma max}})_{\rm{Q}}}\rangle$ in the case where the reaction goes in several channels at a fixed $E_{\rm{\gamma max}}$ value, with production of one and the same final product-nucleus. So, the $^{27}\!\rm{Al}(\gamma,\textit{x})^{24}\rm{Na}$  reaction may be regarded as a monitor of the bremsstrahlung $\gamma$-flux in the energy range $E_{\rm{\gamma max}} = 30 \div 100$~MeV, provided that the average cross-section per equivalent quantum is applied. 

On the other hand, the use of $\langle{\sigma(E_{\rm{\gamma max}})}\rangle$ makes it possible to consider in more detail the dependence of the cross section on the bremsstrahlung $\gamma$-quantum energy, since it is insensitive to the low-energy part of the bremsstrahlung spectrum. Therefore, in the study of single-channel reactions it is preferable to use $\langle{\sigma(E_{\rm{\gamma max}})}\rangle$.

\section*{Declaration of competing interest}\addcontentsline{toc}{section}{Declaration of competing interest}
\label{Decl}
The authors declare that they have no known competing financial interests or personal relationships that could have appeared to influence the work reported in this paper.




\begin{thebibliography}{00}

\bibitem{1}
A.V.~Varlamov, V.V.~Varlamov, D.S.~Rudenko, M.E.~Stepanov. Atlas of Giant Dipole Resonances. Parameters and Graphs of Photonuclear Reaction Cross Sections. INDC(NDS)-394, IAEA NDS, Vienna, Austria, 1999.
\bibitem{2}
E.G.~Fuller, H.~Gerstenberg. Photonuclear Data - Abstracts Sheets 1955 - 1982. NBSIR 83-2742. U.S.A. National Bureau of Standards, 1986.
\bibitem{3}
M.B.~Chadwick, P.~Obloinsky, P.E.~Hodgson, G.~Reffo, Phys. Rev. C44 (1991) 814, doi.org/10.1103/PhysRevC.44.814.
\bibitem{4}
B.S.~Ishkhanov and V.N.~Orlin, Phys. At. Nucl. 74 (2011) 19, doi.org/10.1134/S1063778811010054.
\bibitem{5}
V.Di~Napoli, A.M.~Lacerenza, F.~Salvetti, H.G. de Carvalho, J.B. Martins, Lett. Nuovo Cimento 1 (1971) 835.
\bibitem{6}
V.Di~Napoli, D.~Margadonna, F.~Salvetti, H.G.~de~Carvalho, J.B.~Martins, Nucl. Instrum. Meth. 93 (1971) 77, doi.org/10.1016/0029-554X(71)90140-6.
\bibitem{7}
A.N.~Gorbunov, F.P.~Denisov, V.A.~Kolotukhin, Sov. Phys. JETP 11 (1960) 783.
\bibitem{8}
R.A.~Meyer, W.B.~Walters, J.P.~Hummel, Nucl. Phys. A122 (1968) 606, doi.org/10.1016/0375-9474(68)90580-0.
\bibitem{9}
V.I.~Noga, Yu.N.~Ranyuk, P.V.~Sorokin, Yad. Fiz. (Sov. Journ. Nucl. Phys.) 19 (1974) 945  (in Russian).
\bibitem{10}
A. Shin, Y.W. Choi, Ji-hun Kim, M. Bae. Development Status of TRACE model for PGSFR Safety Evaluation, Transactions of the Korean Nuclear Society Spring Meeting Jeju, Korea, May, 2014.
\bibitem{11}
B.~Friberg and B.~Forkman, Annual Report 1969, sect. V, A: lc (University of Lund, Lund Institute of Technology, Lund, 1970).
\bibitem{12}
A.~Masaike, J. Phys. Soc. Japan 19 (1964) 427, doi.org/10.1143/JPSJ.19.427.
\bibitem{13}
A.N.~Vodin, O.S.~Deiev, S.N.~Olejnik, Probl. Atom. Sci. Tech. 6 (2019) 122.
\bibitem{14}
A.N.~Vodin, O.S.~Deiev, I.S.~Timchenko, S.N.~Olejnik, A.S.~Kachan, L.P.~Korda, E.L.~Kuplennikov, V.A.~Kushnir, V.V.~Mitrochenko, S.A.~Perezhogin, N.N.~Pilipenko, V.S.~Trubnikov, Probl. Atom. Sci. Tech. 3 (2020) 148.
\bibitem{15}
S.Y.F.~Chu, L.P.~Ekstrom, R.B.~Firestone, The Lund/LBNL, Nuclear Data Search, Version 2.0, February 1999, WWW Table of Radioactive Isotopes, available from http://nucleardata.nuclear.lu.se/toi/.
\bibitem{16}
TALYS -- based evaluated nuclear data library. https://tendl.web.psi.ch/tendl 2019/tendl2019.html.
\bibitem{17}
Electron and Positron Incident. http://geant4.web.cern.ch/geant4/.



\end{thebibliography}
\end{document}